\documentclass[
reprint,
superscriptaddress,
 aps,
]{revtex4-2}
\usepackage[utf8]{inputenc}
\usepackage[T1]{fontenc}
\usepackage{amsmath}
\usepackage{amssymb}
\usepackage{graphicx}
\usepackage[capitalize, nameinlink]{cleveref}
\usepackage{subfigure}
\graphicspath{{img/}}
\usepackage{siunitx}
\usepackage{geometry}
\usepackage{fullpage}
\usepackage{color}

\usepackage{lineno}

\begin{document}

\title{Megahertz-rate Ultrafast X-ray Scattering and Holographic Imaging at the European XFEL}

\author{Nanna Zhou Hagström}
\affiliation{Department of Physics, Stockholm University, 106 91 Stockholm, Sweden}

\author{Michael Schneider}
\affiliation{Max Born Institute for Nonlinear Optics and Short Pulse Spectroscopy, Berlin, Germany}

\author{Nico Kerber}
\affiliation{Institute of Physics, Johannes Gutenberg-University Mainz, 55099 Mainz, Germany}

\author{Alexander Yaroslavtsev}
\affiliation{European XFEL, Holzkoppel 4, 22869 Schenefeld, Germany}
\affiliation{Department of Physics and Astronomy, Uppsala University, Uppsala, Sweden}

\author{Erick Burgos Parra}
\affiliation{Synchrotron SOLEIL, Saint-Aubin, Boite Postale 48, 91192, Gif-sur-Yvette Cedex, France}
\affiliation{Unité Mixte de Physique, CNRS, Thales, Université Paris-Saclay, 91767, Palaiseau, France}

\author{Marijan Beg} 
\affiliation{European XFEL, Holzkoppel 4, 22869 Schenefeld, Germany}
\affiliation{Department of Earth Science and Engineering, Imperial College London, SW7 2AZ London, UK}

\author{Martin Lang}
\affiliation{Faculty of Engineering and Physical Sciences, University of Southampton, Southampton SO17 1BJ, United Kingdom}
\affiliation{Max Planck Institute for the Structure and Dynamics of Matter, Luruper Chaussee 149, 22761 Hamburg, Germany}

\author{Christian M. Günther}
\affiliation{Max Born Institute for Nonlinear Optics and Short Pulse Spectroscopy, Berlin, Germany}
\affiliation{Technische Universität Berlin, Center for Electron Microscopy (ZELMI), Berlin, Germany}

\author{Boris Seng}
\affiliation{Institute of Physics, Johannes Gutenberg-University Mainz, 55099 Mainz, Germany}
\affiliation{Institut Jean Lamour, Nancy, France}

\author{Fabian Kammerbauer}
\affiliation{Institute of Physics, Johannes Gutenberg-University Mainz, 55099 Mainz, Germany}
% ...

\begin{abstract}
The advent of X-ray free-electron lasers (XFELs) has revolutionized fundamental science, from atomic to condensed matter physics, from chemistry to biology, giving researchers access to X-rays with unprecedented brightness, coherence, and pulse duration. All XFEL facilities built until recently provided X-ray pulses at a relatively low repetition rate, with limited data statistics. Here, we present the results from the first megahertz repetition rate X-ray scattering experiments at the Spectroscopy and Coherent Scattering (SCS) instrument of the European XFEL. We illustrate the experimental capabilities that the SCS instrument offers, resulting from the operation at MHz repetition rates and the availability of the novel DSSC 2D imaging detector. Time-resolved magnetic X-ray scattering and holographic imaging experiments in solid state samples were chosen as representative, providing an ideal test-bed for operation at megahertz rates. Our results are relevant and applicable to any other non-destructive XFEL experiments in the soft X-ray range.
\end{abstract}

\maketitle

X-rays have long been used as an advanced characterization tool of matter. They are typically used for diffraction, spectroscopy and imaging experiments with high spatial and energy resolutions. These properties have now been exploited for more than a century to achieve a deep understanding of molecules, solid materials and biological samples, fundamental to the progress of science. The appearance, one decade ago, of X-ray free-electron lasers (XFELs) providing intense X-ray pulses with a high degree of transverse spatial coherence and ultrashort pulses, has opened great opportunities for imaging and time-resolved experiments in atomic physics, condensed matter, chemistry, and life sciences beyond what is possible at synchrotron light sources \cite{ayvazyan2006first, emma2010first, ishikawa2012compact, altarelli2011european, grunbein2018megahertz, allaria2012highly, patterson2010coherent, yun2019coherence, halavanau2019very, pellegrini2016xray}.

XFEL technology constantly advances, particularly in terms of spectral brightness. The European XFEL (EuXFEL) is the first facility able to deliver soft and hard X-ray pulses at megahertz repetition rate generated via a self-amplified spontaneous emission (SASE) process \cite{decking2020euxfel}. This greatly improves the statistics of the collected data and in turn the achievable signal-to-noise ratio within a typical experiment time. While in serial femtosecond X-ray crystallography many copies of the samples can be injected into the beam at megahertz repetition rates for accumulation of data \cite{chapman2011femtosecond}, it remains a challenge to recover or to replenish the sample for condensed matter studies in fields such as magnetism, strongly correlated materials and quantum science. 

In this work, we demonstrate non-destructive, stroboscopic soft X-ray scattering and holography experiments at megahertz repetition rates at the Spectroscopy and Coherent Scattering (SCS) beamline at the EuXFEL, exploiting the opportunities offered by the newly commissioned, custom-made two-dimensional detector able to match the EuXFEL megahertz operation. We illustrate the initial capabilities of the beamline at the time of the presented experiments with representative examples of magnetic scattering and imaging experiments of the type performed at other FELs \cite{vodungbo2012laserinduced, pfau2012ultrafast, graves2013nanoscale,henighan2016generation, buttner2017fieldfree,reid2018phenomenological, dornes2019ultrafast,malvestuto2018ultrafast,weder2020transient,buttner2021observation}. We also estimate the heat load on the sample in these experiments, providing a figure-of-merit to find the optimal experimental parameters.

\section*{Results}
\subsection{Operation of the megahertz-rate beamline and detector}
\begin{figure*}[t]
\begin{center}
\includegraphics[width=0.9\textwidth]{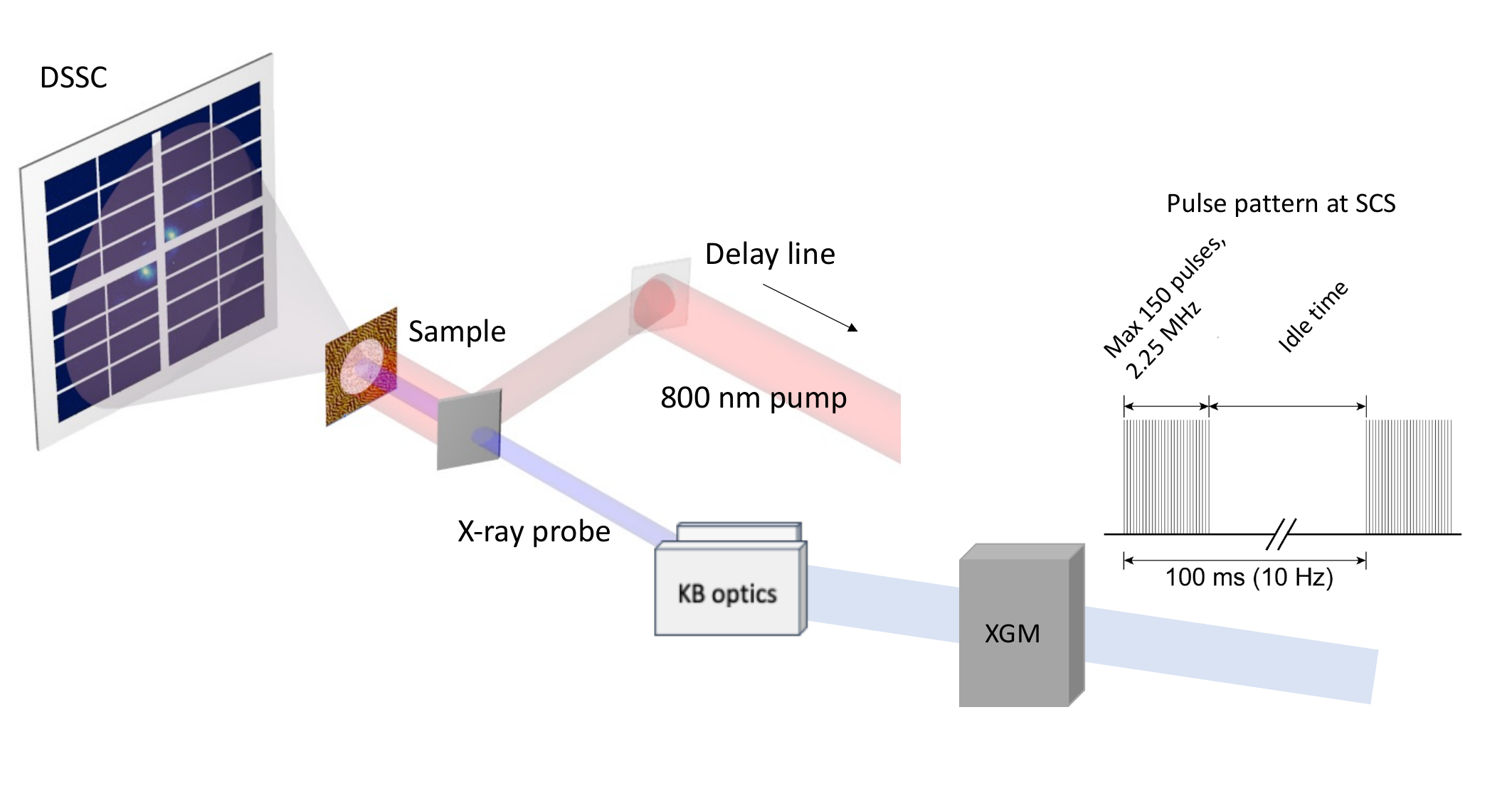}
\caption{\textbf{Schematic of the SCS beamline and of the X-ray pulse structure at the EuXFEL.} The X-ray beam propagates from the right to the left side. The X-rays bursts arrive in trains which contain a user-defined number of pulses. The X-ray gas monitor (XGM) measures the pulse intensity $I_0$ before the focusing KB optics. The pump laser is delivered into the experimental chamber via an auxiliary window and directed to the sample almost parallel with respect to the X-ray beam. The photons scattered by the sample are recorded on the DSSC detector.}
\label{fig:beamline}
\end{center}
\end{figure*}

At the EuXFEL, X-rays arrive in 10~Hz trains of multiple pulses. At the time of the experiment, the number of pulses within a train could be arbitrarily chosen between 1 and 150 separated by at least \SI{440}{\nano \second}, i.e. at a maximum repetition rate of 2.25~MHz within the train, see \cref{fig:beamline}. The SCS beamline covers an energy range of \SIrange{0.25}{3}{\kilo\eV} well suited for core level spectroscopy at the L edges of 3d transition metal (including the most common ferromagnets), the M edges of rare earth elements, and the K edges of lighter elements such as carbon and oxygen. A soft x-ray monochromator provides an energy resolution of approximately \SI{250}{\milli \eV} for the Co and Fe absorption L edges reported in this work ($E/\Delta E\approx3000$), and reduces the pulse energy to tens of microjoules. The pulse duration of the monochromatic X-rays beam is \SI{30}{\fs} on average.

As shown in \cref{fig:beamline}, the incoming intensity $I_0$ of each pulse is monitored by an X-ray gas monitor (XGM)\cite{Maltezopoulos2019xgm}. The beam size at the sample position can be tuned using Kirkpatrick-Baez (KB) mirrors, with a minimal spot diameter of approximately $\SI{1}{\um}$ in both horizontal and vertical directions. Samples are mounted in the forward-scattering fixed target (FFT) chamber, which also includes an electromagnet that can be used to apply magnetic fields of up to \SI{350}{\milli\tesla} parallel to the X-ray beam direction. 
The SCS instrument is equipped with the novel DSSC detector, which can be mounted at different distances from the sample chamber (\SIrange{0.35}{5.4}{\m}), allowing users to cover different scattering wavevector ranges. A multichannel plate-based transmission intensity monitor (not shown in Fig.~\ref{fig:beamline})
simultaneously collects the direct beam after the DSSC detector and is used  to measure the sample absorption. The pump laser beam is inserted in the FFT experiment station with an in-coupling mirror and impinges on the sample nearly collinearly with the X-rays. The laser used here is an YAG-white-light-seeded, non-collinear optical parametric amplifier developed in-house at the EuXFEL providing pump pulses of \SI{800}{\nano \meter} with a duration down of \SI{35}{\femto \second}, which can match the pulse pattern of the XFEL \cite{Pergament2016, palmer2019pump}. The incoming energy can be adjusted from \SI{0.05}{\mJ} up to \SI{2}{\mJ} per pulse with a spot size approximately 50 $\mu$m in diameter. In this work, the sample is always pumped at half the probe repetition frequency in order to obtain pairs of pumped and unpumped measurements that are close in time. This allows users to remove the effect of long-term drift on the measurements.

\begin{figure*}[t]
\begin{center}
\includegraphics[width=0.9\textwidth]{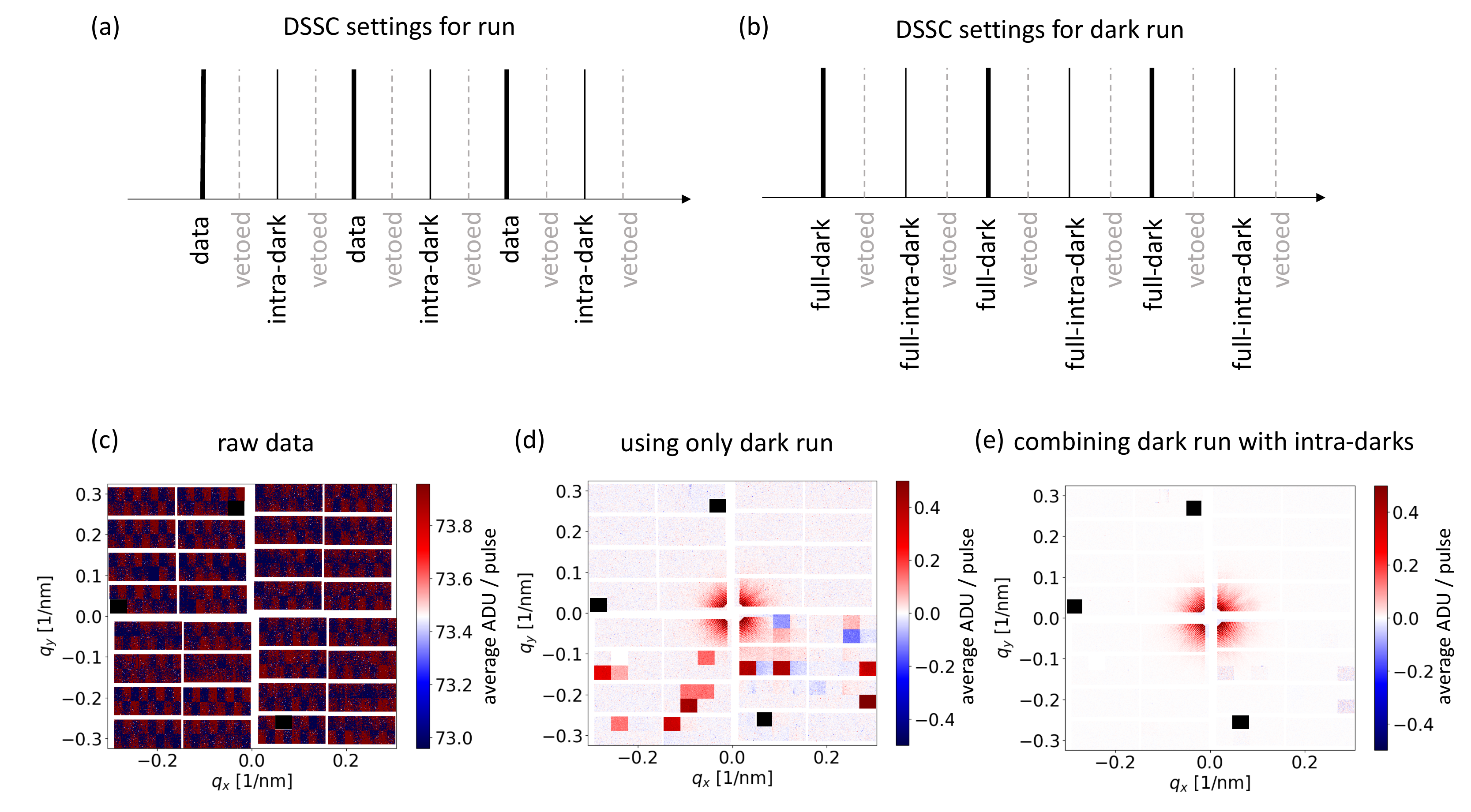}
\caption{\textbf{Schematic of the pulse labeling for the dark subtraction and application example.} X-ray pulse labeling for (a) acquisition with X-rays and (b) without X-rays, a so-called \emph{dark run}. Separate dark runs are usually 1 min for practical reasons (here 90 000 frames). (c) Raw data collected by the DSSC detector plotted around its mean value. (d) Dark subtraction using only a separate dark run and (e) dark subtraction combining a separate dark run and the intra-dark events.}
\label{fig:darksub}
\end{center}
\end{figure*}

The DSSC detector is presently the fastest 1-megapixel camera available worldwide, providing single-photon sensitivity in the soft X-ray regime. It is capable of recording data from the full pixel array with a \SI{220}{\nano\second} frame interval, corresponding to a \SI{4.5}{\mega\hertz} repetition rate. The data is retrieved in the 10~ms long inter-pulse train gap of the FEL. The sensitive area of the camera is about 505~\si{\centi\meter ^2} in size, composed of $1024 \times 1024$ equilateral hexagonal pixels with a side length of \SI{136}{\micro\meter}. The present camera uses for each hexagonal pixel a miniaturised silicon drift detector (MiniSDD) coupled to a linear readout electronics front-end. The camera comprises 16 sub-units called "ladders" (horizontal blocks) arranged into four quadrants. Each ladder has 2 monolithic sensors and is read out by 16 independent readout application specific integrated circuit (ASICs) \cite{porro2021minisddbased}. The four quadrants can be moved independently if required by the experiment, while the location of the ladders within one quadrant is fixed.

While the DSSC detector always runs at 4.5~MHz, a ``veto'' system allows discarding frames according to a user-defined pattern or an additional signal provided by an external veto source. When pulses are delivered at a smaller frequency than 4.5~MHz, the user can choose to record frames at the same frequency as the FEL, or at a higher one, in order to collect so-called \emph{intra-dark} frames in-between data frames, see \cref{fig:darksub}(a)-(b). Discarding (vetoeing) of unused frames is crucial to minimize the amount of data collected and perform efficient analysis. In fact, at full repetition rate, the camera produces data at a rate of 134~Gbit/s, which would lead to single experiments creating petabytes of data. \cref{fig:darksub}(c) is an example of the raw data collected by the DSSC detector, the uncorrected image has a mean of 73.35 ADUs, which is almost entirely an offset signal due to the analog-to-digital converters \cite{hansen20138bit, porro2021minisddbased}, which can be removed by appropriate signal subtraction.

The first dark signal subtraction, pixel-by-pixel, is made using dark frames acquired in a separate run with the same settings of the DSSC camera (gain and veto pattern), but without X-rays hitting the detector. This is labeled as a \emph{dark run}, and subtraction of such a run from the data results in the plot in \cref{fig:darksub}(d). The few darker squares in the figure are due to the fact that for a few random frames, the ASICs did not transfer the acquired data correctly. This is due to a firmware bug that was solved after the experiment. A separate dark run helps removing the large static electronic offset, but does not correct for other sources of noise, such as the signal generated back-scattered photons or other systematic electronic effects which are occurring during the measurements. These can however be removed using the intra-darks signal, closer in time to the signal events. By combining the dark run with the intra-darks, one can achieve the most appropriate background subtraction, as shown in Fig.~\ref{fig:darksub}(e), where the image was calculated as 
\begin{center}
[\text{run(data frame $-$ intra-dark frame})] $-$ [\text{dark run(data frame $-$ intra-dark frame})].
\end{center}
Note that three black squares indicate ASICs that were damaged and cannot be used for data collection \cite{porro2021minisddbased}. 
We estimate an experimental root-mean-square (RMS) noise for each pixel $\sigma/\sqrt{N}\approx5\cdot10^{-3}$ ADUs, where $\sigma\approx1.4$ ADUs is the standard deviation and $N\sim10^5$ is the number of events in a measurement run. With the four data sets needed for complete offset subtraction, this leads to a total RMS noise $\sigma_{\rm tot}/\sqrt{N}\approx10^{-2}$ ADUs, which allows to readily measure signals in the 0.1 - 1 ADU range, as shown in Fig. \ref{fig:darksub}(e).

\subsection{Ultrafast small-angle X-ray scattering at megahertz repetition rates}
\begin{figure*}[t]
\begin{center}
\includegraphics[width=\textwidth]{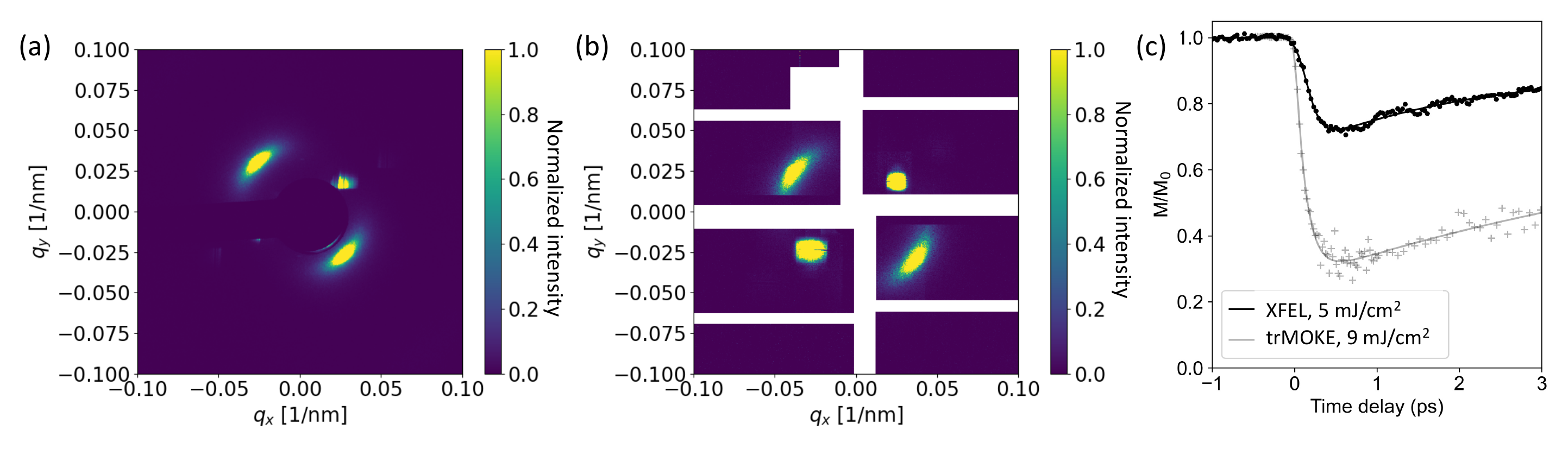}
\caption{\textbf{Megahertz-rate time-resolved magnetic X-ray scattering.} Resonant Co L$_3$ edge scattering pattern of a CoFe/Ni thin film multilayer recorded at (a) the SEXTANTS beamline at SOLEIL synchrotron and (b) at the SCS beamline at the EuXFEL. The first order magnetic scattering is observed along the top-left to bottom-right diagonal. The scattering from the non-magnetic grating is the feature visible along the opposite diagonal. The intensity is in linear scale and normalized to the maximum magnetic scattering amplitude. (c) Time-resolved pump-probe data recorded on the same sample. Black symbols: data from the EuXFEL, computed as the average intensity of the first order peak in the frames when the pump laser was impinging on the sample, divided by the nearest previous unpumped frame. Gray symbols: data from a table-top MOKE setup with different pump fluence. The solid lines show the fit to the data. Further details are given in the main text}
\label{fig:SAXS_trSAXS}
\end{center}
\end{figure*}

Small-angle X-ray scattering (SAXS) in the soft X-ray regime has been shown to be a unique tool to explore not only the temporal, but also the spatial dynamics of ultrafast processes on nanometer length scales. In ultrafast magnetism, this capability has been proven to be a crucial feature, since many of the fundamental physical processes at play are strongly connected to the nanometer structure in the material~\cite{vodungbo2012laserinduced, pfau2012ultrafast, graves2013nanoscale, bergeard2015irreversible, iacocca2019spincurrentmediated, hennes2020laserinduced}. We measure CoFe/Ni multilayer samples with out-of-plane magnetization showing ordered stripe domains with a typical domain size in the range of \SIrange{115}{125}{\nano \meter}, as revealed by magnetic force microscopy (MFM), see SI. Due to the XMCD effect, the magnetic stripe domains act as an absorption grating for linearly polarized photons in resonance with the Co L$_3$ absorption resonance at approximately 778 eV \cite{hellwig2003xray}. This gives rise to an anisotropic scattering signal along a preferential axis. The sample also comprises a curved diffraction grating milled in the silicon nitride carrier membrane, creating a non-resonant reference scattering signal on the detector~\cite{schneider2016curved}. 
The DSSC camera is placed \SI{2}{\m} from the sample and the X-ray beamsize is \SI{75}{\micro \meter}. As optical pump, we use \SI{800}{\nano \meter}, \SI{100}{\femto \second} laser pulses. The pump laser is operated at a repetition rate of \SI{282}{\kHz} with 10 pulses per train, while the XFEL runs at \SI{564}{\kHz} with 20 pulses per train, allowing to record unpumped X-ray scattering frames in-between pumped ones. 


A typical scattering pattern from the magnetic stripe domains recorded from the SEXTANTS beamline at the SOLEIL synchrotron \cite{sacchi2013sextants} is shown in \cref{fig:SAXS_trSAXS}(a), with the corresponding XFEL data in \cref{fig:SAXS_trSAXS}(b). In both images, we observe two broad features arising from the scattering of X-rays from the magnetic domains along the top-left/bottom-right diagonal of the image, as well as the smaller features related to the reference diffraction grating along the opposite diagonal. The synchrotron image is acquired with an average photon rate of $5\cdot10^{12}$~photons/s and \SI{1}{\second} exposure time while for the XFEL data a total of $9\times10^{11}$~photons were incident on the sample, with 50 pulses per train and 600 trains in total with an average of $3\times10^{10}$~photons/pulse. 

The black symbols in \cref{fig:SAXS_trSAXS}(c) show the laser induced ultrafast dynamics  of the magnetic scattering spot intensities, measured in a pump-probe configuration, with a pump fluence of \SI{5}{\mJ/\cm^2} and with the sample at magnetic remanence. In the same plot, we compare the XFEL data with the one recorded on the very same sample using a table-top time-resolved magneto-optical Kerr effect (tr-MOKE) setup with a saturating magnetic field and with a pump fluence of \SI{9}{\mJ/\cm^2}. Both curves describe the laser induced ultrafast demagnetization of the ferromagnetic film \cite{beaurepaire1996ultrafast}. The curves were fitted using the formula derived from a three-temperature model \cite{beaurepaire1996ultrafast, malinowski2008control, hennes2020laserinduced}, i.e. $M(t)=1-(A-Be^{-t/\tau_M}-Ce^{-t/\tau_R})\circledast\Gamma(t)$, where $\tau_M$ is the demagnetization time and $\tau_R$ is the picosecond recovery time, different from the thermal one with much larger time constant. The constants $A$, $B$, and $C$ are amplitudes that can be related to the different physical processes. Here we are only interested in the time constants, and we neglect further considerations on these amplitudes. The convolution with a Gaussian function $\Gamma(t)$ takes into consideration the finite pulse durations which were different for the tr-SAXS and tr-MOKE measurements, and allows us to extract the true demagnetization constant. From the fit of the XFEL data, we find $\tau_m=102\pm8$ fs and $\tau_R=2.18\pm0.07$ ps, while from the tr-MOKE we obtain $\tau_m=129\pm10$ fs and $\tau_R=6.08\pm0.5$ ps. The slightly smaller time constants retrieved for the XFEL measurements are consistent with a smaller quenching of the sample \cite{koopmans2010explaining}.

\begin{figure*}[t]
\begin{center}
\includegraphics[width=0.9\textwidth]{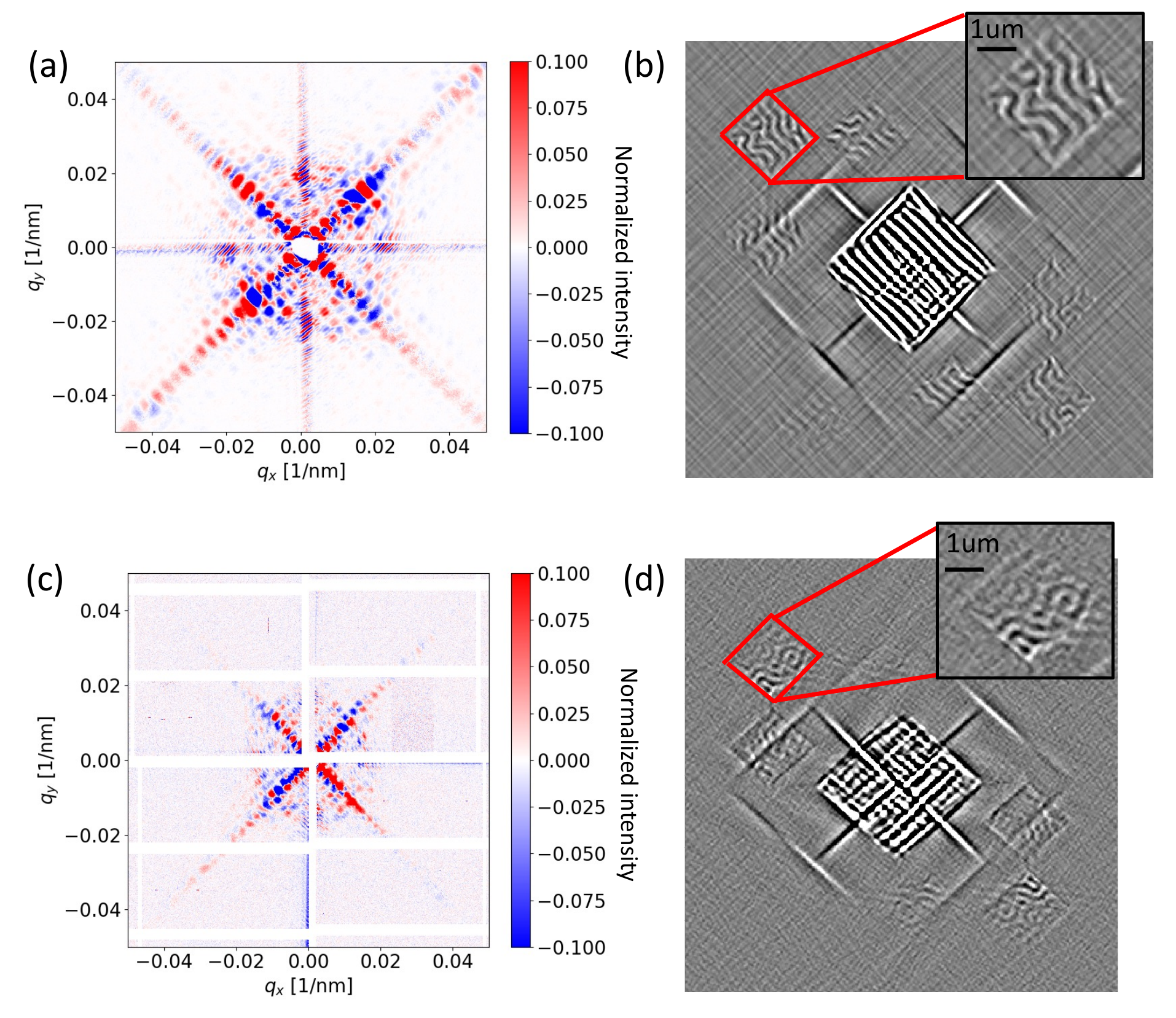}
\caption{\textbf{Megahertz-rate magnetic X-ray holographic imaging.}  Magnetic hologram of a CoFeB thin film multilayer recorded at the Fe L$_3$ edge at (a) the COMET endstation at the SEXTANTS beamline at SOLEIL synchrotron and at (c) the SCS beamline at the EuXFEL at a 2.25~MHz repetition rate. The intensity is in linear scale and normalized to the maximum intensity value. Reconstructions of the magnetic domains using the HERALDO technique on (b) the synchrotron data and (d) the free-electron laser one.}
\label{fig:imaging}
\end{center}
\end{figure*}

\subsection{X-ray holographic imaging at megahertz repetition rates}
High-resolution X-ray imaging techniques are mostly of two kinds: those based on Fresnel-type optics, and those which are lensless. While the former type has found much application at synchrotron lightsources, they are difficult to realise in the soft X-ray region at free-electron lasers due to the risk of damage by strong absorption of intense X-ray pulses. In these facilities, lensless techniques are preferred for full field imaging, since they can exploit the high degree of transverse coherence of FEL radiation \cite{wang2012femtosecond, vonkorffschmising2014imaging, willems2017multicolor}. X-ray holography is one such lensless imaging technique that relies on the interference between two beams, where one holds information about the sample, and the other acts as the phase reference. A Fourier transform of the two-dimensional diffraction reconstructs the real-space image. The samples are thin CoFeB multilayer films with out-of-plane magnetization. From magnetic force microscopy (MFM) we observe approximately \SI{200}{\nano\m} wide labyrinth magnetic domains at remanence. The holography aperture is a square with a side of \SI{2.5}{\micro\m}, rotated by \SI{45}{\degree} with respect to the sides of the X-ray transparent window where the film is deposited. The reference beam is generated by two orthogonal slits in the holography mask (see SI for details). This allows to reconstruct the image using the HERALDO technique \cite{zhu2010highresolution, duckworth2011magnetic}, which mitigates the artefacts due to the detector gaps. The sample was pre-characterized at the COMET endstation at the SOLEIL synchrotron \cite{popescu2019comet}. In \cref{fig:imaging}(a) we plot the magnetic scattering signal recorded at the synchrotron, calculated as the difference between the signal taken with X-rays of opposite helicities at the Fe L$_3$-edge, i.e. at approximately \SI{707}{\eV}. In \cref{fig:imaging}(b), we show the corresponding image reconstruction applying the full HERALDO procedure \cite{zhu2010highresolution, duckworth2011magnetic}. The image reveals the presence of magnetic domains in one of the six smaller squares which are the cross-correlation between the object and the three corners of the L-shaped reference slit. Each corner yields a pair of conjugated images, where the opposite contrast indicates oppositely oriented magnetic domains.
The XFEL measurement on the same sample - with different magnetic domain pattern due to exposure to a magnetic field between the respective measurements - is shown in \cref{fig:imaging}(c), where in this case the X-ray helical polarization at the required photon energy  is achieved with a thin Fe film polarizer inserted in the beam before the sample~\cite{muller2018note}, at the expense of photon flux. Helicity reversal is obtained by reversing the magnetic field applied to the thin film polarizer. The detector is placed \SI{4.6}{\m} away from the sample, in order to record the magnetic information in the lower $q$-range. The beam spot is \SI{50}{\micro \meter} in diameter, smaller than in the case of the SAXS experiment, but much larger than the holography apertures. The samples are probed with different repetition rates of the XFEL between 0.226~MHz and 2.25~MHz with no sample damage observed. This can be partly explained by the thick gold layer where the holography mask is patterned, as we discuss in details in the last part of this work. The hologram is the result of 15 min acquisition (1000 pulses/s) and $4 \times 10^{13}$~photons on the sample area for each helicity. As a comparison, the photon count on the same HERALDO FTH sample area at the COMET end-station of the SEXTANTS beamline was $1\times10^{13}$~photons acquired in 90~s. 
\cref{fig:imaging}(d) shows the 2D Fourier transform of the hologram of the XFEL data. Like in \cref{fig:imaging}(b) we observe the auto-correlation of the object aperture in the center of the image, and three pairs of reconstructions.


\begin{figure*}[t!]
\begin{center}
\includegraphics[width=\columnwidth]{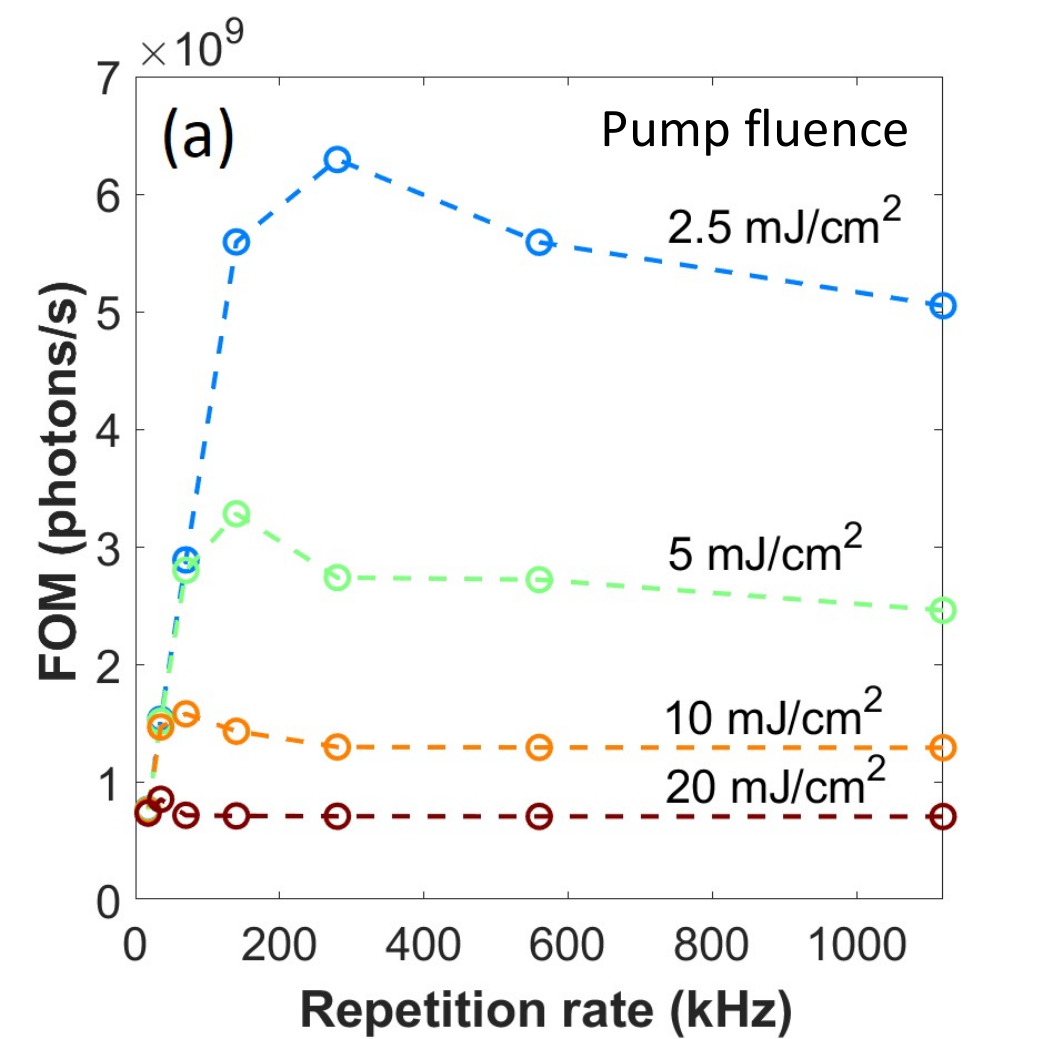} \includegraphics[width=\columnwidth]{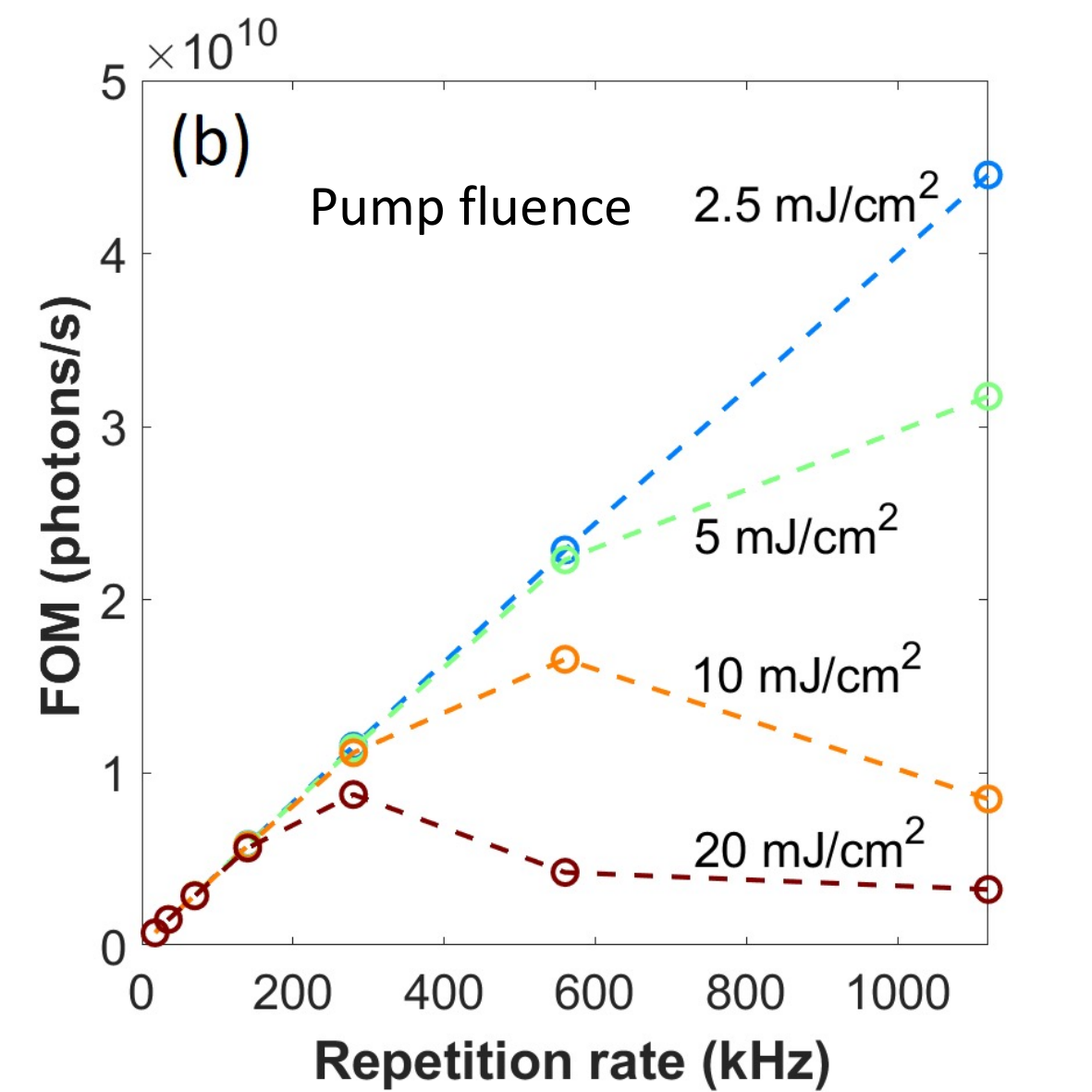}
\caption{\textbf{Figure of merit for magnetic scattering as a function of repetition rate of XFEL pulses.} Simulated photons per second of magnetic scattering as a function of XFEL repetition rate in a burst of \SI{280}{\micro \second} and for different pump fluences: (a) bare thin film samples on Si membranes and (b) samples on Si membranes with an additional \SI{500}{\nano \meter} heat sink layer. Note that the FOM scales are different in (a) and (b) by a factor of 10.} 
\label{fig:FOM}
\end{center}
\end{figure*}

\section*{Discussion}

When comparing the SAXS measurements in Fig.~\ref{fig:SAXS_trSAXS}, we note that the number of pulses per train had to be reduced to 50 in order to keep the sample unchanged by the X-rays, subsequently the average photon flux (photons/s) is 2 orders of magnitude smaller compared to the one of the synchrotron, mostly limited by the burst mode operation of the machine. Naturally, the XFEL measurements are performed using femtosecond X-ray pulses, which allows for ultrafast experiments that are not feasible at a synchrotron. We have also confirmed that the extracted time constants with table-top and XFEL experiments are comparable, demonstrating the reliability of the XFEL measurements in measuring ultrafast dynamics.

Looking at the holographic imaging data, we notice that despite the fact that the XFEL image is slightly noisier, the magnetic domains are clearly distinguishable. We believe that part of the issue is also a non-ideal illumination of the holographic mask which can be readily improved with an optimized design. Nevertheless, these data demonstrate that a full magnetic image reconstruction at the EuXFEL is possible within tens of minutes. Hence, ``movies'' of the magnetization combining ultrafast and nanometer resolutions are now possible at a free-electron laser within a typical beamtime allocation. Future upgrade of the instrument, such as circularly polarized photons directly generated by the undulators, will likely be able to shorten the acquisition time by up to one order of magnitude. The gain is expected to mostly arrive from the greatly reduced correlation between intensity and polarization caused by a thin film polarizer, which cannot be easily normalized out.

Finally, we estimate the possible heating effects of X-rays pulses at high repetition rate on the samples. We perform heat diffusion simulations, and we use the dependence of the magnetization on temperature to calculate the loss of signal due to heat. The details are given in the Methods section. These calculations allow us to calculated a figure of merit (FOM) which can then be plotted as a function of XFEL repetition rate (considering the actual pulse structure), and for different pump fluences, as shown in Fig. \ref{fig:FOM}(a)-(b). The FOM is determined by the competition of two processes: the number of photons reaching the detector, which increase linearly with the average X-ray power, and the amount of meaningful signal (proportional to the magnetization squared), which decreases with average power. Thus, the FOM can be interpreted as the number of information-carrying photons hitting the detector over a given time. We find that the optimal repetition rate is in the order of 100~kHz for pump-probe measurements on typical samples on free-standing membranes, which can be pushed to the megahertz rate if a proper heat sink layer is implemented within the sample, such as for the case of holographic imaging experiments.

\section*{Methods}
\footnotesize
\noindent\textbf{Sample preparation}
The CoFe/Ni thin film multilayers with a composition of Ta(\SI{3}{\nm})/Cu(\SI{5}{\nm})/[CoFe(\SI{0.25}{\nm})-Ni(\SI{0.75}{\nm})]$_{20}$/CoFe(\SI{0.25}{\nm})/Cu(\SI{3}{\nm})/Ta(\SI{3}{\nm}) were deposited on \SI{200}{\nano \meter} thick Si membranes with a lateral size of 2~mm. Sample thicknesses were calibrated with X-ray reflectometry. The diffraction grating in the Si$_3$N$_4$ membrane was fabricated using a focused Ga$^+$ ion beam (FIB) system. The magnetic domains were aligned to stripes after in-plane demagnetization and were characterized via SAXS at the VEKMAG endstation at the BESSY II synchrotron \cite{noll2017vekmag} and at the RESOXS endstation of the SEXTANTS beamline at SOLEIL synchrotron, as well as by magnetic force microscopy.  The X-ray holography samples were magnetic multilayer films [Ta(\SI{5}{\nm})/Co$_{20}$Fe$_{60}$B$_{20}$(\SI{0.9}{\nm})/MgO(\SI{2}{\nm})]$_{15}$ with out-of-plane magnetization, were produced by DC magnetron sputtering deposition. The material was deposited on Si$_3$N$_4$ membranes. The HERALDO holography mask was fabricated by milling reference through the \SI{1}{\micro\m} thick Au layer using a FIB system. The reference slits (\SI{40}{\nano\meter} wide and \SI{4}{\micro \meter} long) are milled through the Au, the Si$_3$N$_4$ membrane and the magnetic thin film while only the Au is removed over the sample (object hole). The samples were characterized at the COMET endstation at SEXTANTS beamline at SOLEIL synchrotron as well as by magnetic force microscopy.\\

\noindent\textbf{Data collection and analysis} 
During the beamtime, more than 780~terabytes of data were captured using the EuXFEL's control and acquisition system~\cite{Karabo2019}. Offline data analysis was directed from Python and Jupyter notebooks~\cite{jupyterxfel2019}, making use of the storage, calibration, compute, and data analysis infrastructure at EuXFEL~\cite{kuster2014detectors, Fangohr2017}. Analysis tools that were developed for this work and that can be re-used for similar research, have been integrated into the EuXFEL open source software data analysis stack~\cite{software-euxfel2021}.\\

\noindent\textbf{Heat diffusion simulation}
The fraction of X-ray and optical pulse energy absorbed by the CoFe/Ni multilayered sample was calculated using the optical constants and refractive indexes of the sample materials for certain photon energies, available from online databases. The subsequent heat diffusion in the layers was simulated with the equation
\begin{equation}
{\rho}C\frac{dT_n}{dt}-k{\Delta}T_n+h(T_n-T_{n+1})=0.
\label{eq:heatdif}
\end{equation}
The first and second terms of \cref{eq:heatdif} describe the heat diffusion in the layer $n$, while the third term introduces the heat exchange between the layers $n$ and $n+1$. In \cref{eq:heatdif}, $T$ is the temperature of the layer $n$, $\rho$ is the mass density, $C$ is the heat capacity and $k$ is the thermal conductivity of the respective layer, $h$ is the coefficient of heat transfer between layers $n$ and $n+1$. The $h$ value depends on the thermal conductivity and the thickness of two layers, as well as the thermal conductance of the interface between them \cite{lyeo2006thermal}. \cref{eq:heatdif} was solved numerically for each layer of the sample in the polar coordinates. We assumed that the system is two-dimensional since the thickness of the layers is much smaller than their lateral size. In the heat diffusion simulation, the lateral sample size was \SI{2}{\milli\m}, the spacing for the computation grid was \SI{2}{\micro\m}, the total time of the simulation \SI{270}{\micro\s} and the time step \SI{1.35}{\pico\s}. While varying the X-ray and pump pulses repetition rate, the pump was always kept at half the frequency of the X-ray probe. We used the constant temperature boundary conditions, assuming the perfect heat removal from the sample by the perimeter, which is always maintained at room temperature. All parameters of the simulation were taken as constants at room temperature. The magnetization was estimated from the temperature values using the mean field approximation \cite{kittel1996introduction}:
\begin{equation}
M={M_S}\tanh\left(\frac{T_C}{T}\frac{M}{M_S}\right),
\label{eq:meanfield}
\end{equation}
where $T$ and $M$ are the average temperature and magnetization of the magnetic layers within the X-ray beam spot, $T_C=750$ K is the Curie temperature of the CoFe/Ni sample and $M_S=1$ is the saturation magnetization.

\normalsize

\section*{Acknowledgments}
The authors acknowledge European XFEL in Schenefeld, Germany, for provision of X-ray free-electron laser beamtime at Scientific Instrument SCS and thank the instrument group and facility staff for their assistance. The authors from the DSSC consortium want to thank all engineers, technicians, postdocs and students who have contributed to the design, the development and the assembly of the camera.
The authors would like to thank the teams of the SEXTANTS beamline at SOLEIL synchrotron (proposal ID 20160880 for the characterization of static properties of the FeCo/Ni multilayers, and through in-house beamtime for the static holography) and the VEKMAG endstation at BESSY II synchrotron for the static characterization of the samples. E.J. is grateful for the financial support received from the CNRS-MOMENTUM. A.Y. acknowledges support from the Carl Trygger Foundation. L.M., M.R., A.P.K., W.R., and G.G are funded by the Deutsche Forschungsgemeinschaft (DFG, German Research Foundation) – SFB-925 – project 170620586. R.J. and R.K. acknowledge support from AFOSR Grant. No. FA9550-19-1-0019. N.K., B.S., F.K., M.Kläui acknowledge support by the DFG (SFB TRR 173 Spin+X 268565370) and Topdyn. M.Baidakova acknowledges support from Ministry of Science and Higher Education of the Russian Federation (agreement № 075-15-2021-1349). N.Z.H., M.P., K.N, D.P., and S.B. acknowledge support from the European Research Council, Starting Grant 715452 MAGNETIC-SPEED-LIMIT.

\section*{Author Contributions}

N.Z.H. coordinated the data analysis of both experiments, with contributions from M.Schneider, E.B.P., M.B., A.Y., E.I., N.J., A.Scherz, S.B., and E.J. N.Z.H, E.B.P and N.J. calculated the holography reconstructions. M.Schneider developed the DSSC data analysis tools. J.M.S fabricated the SAXS samples. N.K., B.S, F.K., and M.Kläui fabricated the holography samples. M.Schneider, C.M.G., and S.E prepared the holography masks and absorption gratings. A.Y. performed the heat diffusion simulations. F.R., K.C., C.L., A.P.K, E.J., L.Müller, N.Z.H., N.K., S.O. performed the experiment at VEKMAG. H.P., E.B.P., E.J., and N.J. performed the SOLEIL experiments and corresponding data analysis. A.Scherz coordinated the experiment at the SCS instrument at the European XFEL. J.T.D., C.Boers, A.R. participated to the experiment preparation and setup. D.H., R.C., J.Schlappa, B.V.K., R.G., L.Mercardier, N.A., L.L.G., M.Teichmann, A.Y., G.M., and A.Scherz operated the SCS instrument. S.H., D.B., J.Szuba, and K.W. developed the data acquisition and management system to operate with the DSSC data rates. M.Lang, M.Beg, R.R., and H.F. designed and implemented tailored data analysis software. T.K., M.Bergemann, E.K., M.Spirzewski, and H.F. developed data analysis software to read and process the data of the new DSSC detector at SCS. J.Z. provided tailored software for the online data analysis. M.Porro lead the DSSC project. A.C., C.F., P.F., K.H., M.M. and C.B.W. contributed and coordinated the electronics and mechanics development, production, commissioning and calibration of the DSSC camera. A.C., F.E., S.M. and M.Porro contributed to detector optimization during the experiment. M.Kuster lead the detector activities at the European XFEL and contributed to the development of calibration methodology and software, integration and testing of the DSSC detector. J.E., D.L., A.Samartsev, and M.Turcato prepared the DSSC detector for the experiment and supported operation during the experiment. M.Lederer led the laser development, installation and commissioning activities at European XFEL. D.R., J.W., D.Kane, S.V., J.Meier, F.P. and T.J. were the main contributors to installation and commissioning at SASE 3. The sample environment was developed by C.D. and J.Moore. N.Z.H., M.Schneider, N.K., E.B.P., M.Beg, M.Lang, M.Pancaldi, K.N., D.P., R.J., S.B.H., S.K.K.P., S.O., D.T., D.Ksenzov, C.Boeglin, I.P., M.Baidakova, C.v.K.S., B.V., L.Müller, M.M.F., A.P.K., M.R., W.R., G.G., T.K., R.R., E.E.F., O.S., C.G., C.S.H., H.A.D., E.I., H.T.N., J.M.S., M.W.K., T.J.S., R.K., H.F., S.E., M.Kläui, N.J., S.B., and E.J. performed the experiments and live analysis at the SCS instrument. M.Borchert and C.v.K.S. characterized the optical damage of the magnetic samples and performed the optical MOKE experiments. E.J. designed and coordinated the SAXS experiment with the help of B.V., T.J.S., H.T.N., A.P.K., L.Müller, and  R.C.. S.B. designed and coordinated the holography experiment with contribution from N.Z.H., N.K. and A.Scherz. S.B. wrote the manuscript draft, with substantial contribution from N.Z.H, A.Y., N.J., and E.J.. All co-authors contributed to the manuscript.

\section*{Competing Interests statement}

The authors declare no competing interests. 

\section*{Data and materials availability} 
All data are available in the main text or the supplementary materials. Raw data generated at the European XFEL are available at: \\ DOI: 10.22003/XFEL.EU-DATA-002212-00, \\ DOI: 10.22003/XFEL.EU-DATA-002222-00 and \\ DOI: 10.22003/XFEL.EU-DATA-002530-00

\section*{Corresponding authors}
\noindent Correspondence to\\andreas.scherz@xfel.eu,\\ emmanuelle.jal@sorbonne-universite.fr and \\ stefano.bonetti@fysik.su.se

\bibliographystyle{apsrev}
\bibliography{TechnicalArticle.bib}

\end{document}